\documentclass[english,prl,preprint,superscriptaddress]{revtex4}

\usepackage{times}
\usepackage[T1]{fontenc}
\usepackage[latin1]{inputenc}
\usepackage{amsmath}
\usepackage{amssymb}
\usepackage{MnSymbol}
\usepackage{xfrac}
\usepackage{float}
\usepackage{color}
\usepackage{graphicx}
\usepackage{xfrac}
\makeatletter
\usepackage{babel}
\makeatother

\begin{document}

\title{Collapse of the Cooper pair phase coherence length at a superconductor to insulator transition}

\date{\today{}}
\author{S. M. Hollen* (Shawna\_Hollen@brown.edu)}
\affiliation{Department of Physics, Brown University, Providence, RI 02912}
\author{\\G. E. Fernandes (Gustavo\_Fernandes@brown.edu)}
\affiliation{School of Engineering, Brown University, Providence, RI 02912}
\author{\\J. M. Xu (Jimmy\_Xu@brown.edu)}
\affiliation{Department of Physics, Brown University, Providence, RI 02912}
\affiliation{School of Engineering, Brown University, Providence, RI 02912}
\author{\\J. M. Valles, Jr. (James\_Valles\_Jr@brown.edu)}
\affiliation{Department of Physics, Brown University, Providence, RI 02912}

\begin{abstract} 

We present investigations of the superconductor to insulator transition (SIT) of uniform a-Bi films using a technique sensitive to Cooper pair phase coherence.  The films are perforated with a nanohoneycomb array of holes to form a multiply connected geometry and subjected to a perpendicular magnetic field.  Film magnetoresistances on the superconducting side of the SIT oscillate with a period dictated by the superconducting flux quantum and the areal hole density.  The oscillations disappear close to the SIT critical point to leave a monotonically rising magnetoresistance that persists in the insulating phase.  These observations indicate that the Cooper pair phase coherence length, which is infinite in the superconducting phase, collapses to a value less than the interhole spacing at this SIT.  This behavior is inconsistent with the gradual reduction of the phase coherence length expected for a bosonic, phase fluctuation driven SIT.   This result starkly contrasts with previous observations of oscillations persisting in the insulating phase of other films implying that there must be at least two distinct classes of disorder tuned SITs.

\end{abstract}

\maketitle

Superconductor to insulator quantum phase transitions (SIT) can be induced in a wide range of quasi two-dimensional superconducting systems, including elemental films, high T$_\mathrm{c}$ superconductors, organic superconductors and superconductor graphene composites\cite{Gantmakher:PUsp2010,Bollinger:Nature2011, Beschoten:PRL1996, Leng:PRL2011, Fazio:PhysRep2001, Bollinger:PRL2008,Allain:NatMat2012,Shi:NatMat2012}.  Remarkably, these transitions occur, nearly universally, at a critical normal state resistance, $R_\mathrm{Nc}\simeq R_\mathrm{Q} =h/(2e)^2$, and film resistances often show scaling behavior around this critical point.  The most prominent theories that can account for these behaviors view the SIT as a Cooper pair or boson localization transition, rather than a Cooper pair breaking transition\cite{Fisher:PRL1990, Ghosal:PRL1998, Bouadim:NatPhys2011, Dubi:Nature2007, Feigelman:APhys2010, Falco:PRL2010}.   Recent experiments that probe films near the SIT in new ways have provided more details that support this "bosonic" picture of the SIT\cite{Sambandamurthy:book, Sherman:PRL2012, Stewart:Science2007, Sambandamurthy:PRL2004, Sacepe:NatPhys2011}.  For example, scanning tunneling microscopy (STM) has revealed that spatial inhomogeneities develop in the order parameter on approaching the SIT suggesting that Cooper pairs localize into islands\cite{Sacepe:NatPhys2011,Sacepe:NatComm2010}.  High frequency transport measurements indicate that a finite superfluid density persists in non-superconducting films\cite{Crane:PRB2007}.  Here, we describe experiments employing a technique that is uniquely sensitive to the length scale that diverges at a bosonic SIT: the phase coherence length $\xi_\phi$.  This technique previously revealed that Cooper pairs maintain their phase coherence over 100's of nanometers through the SITs of two distinct film systems\cite{Stewart:Science2007, Kopnov:PRL2012}.  For this work, we investigate a third film system\cite{Valles:PRL1992, Chervenak:PRB1999, Haviland:PRL1989} that is likely to provide the most stringent test of whether the bosonic SIT is generic to thin films.  

This phase sensitive technique requires patterning films with an array of nanometer-scale holes.  The hole patterning creates a simply connected geometry that leads to Little-Parks-like (LP) oscillations\cite{LittleParks:PR1964} in the magnetoresistance (MR).  The oscillations occur provided the CP phase coherence length ($\xi_{\phi}$) exceeds the inter-hole spacing ($\sim$100 nm).   For a bosonic SIT, $\xi_{\phi}$ starts at infinity in the superconducting phase and decreases algebraically with a power of order 1 with distance from the critical point\cite{Fisher:PRL1990b,Sondhi:RMP1997}.  Thus, oscillations are expected to persist well into the insulating phase for a bosonic SIT.   By contrast, in the fermionic scenario in which the order parameter amplitude goes to zero at the SIT,  $\xi_{\phi}$ collapses to zero at the critical point. We summarize these opposing pictures of the SIT in Fig. 1. Using this embedded hole array technique, we previously uncovered a CPI phase in a-Bi films deposited on anodized aluminum oxide (AAO) substrates\cite{Stewart:Science2007}.  Later we found that these substrates induced spatial film thickness variations and thus, Cooper pairing inhomogeneities that could give rise to a bosonic SIT\cite{Hollen:PRB2011}.  Here, we present results from similar experiments on a-Bi films fabricated to be uniformly thick by using atomically flat Si substrates.  Little-Parks-like oscillations demonstrating CP phase coherence appear only on the superconducting side of the SIT in these uniform films.  The results are consistent with $\xi_{\phi}$ collapsing at this SIT in accord with a fermionic SIT.   These results imply that there are at least two separate classes of disorder-driven SITs.  

Motivation for these investigations came from developments over the last decade that strongly suggest that the bosonic picture of the SIT is universal\cite{Trivedi:book}.  Most convincingly, there is direct evidence that Cooper pairs (CPs) can remain intact while becoming localized and phase incoherent through the SIT to form a Cooper Pair Insulator (CPI) phase\cite{Sambandamurthy:book, Sherman:PRL2012, Stewart:Science2007, Sambandamurthy:PRL2004, Sacepe:NatPhys2011}.  STM experiments on superconducting TiN and In oxide (InO$_{x}$) films indicated that the pairing energy gap stays finite through the SIT while developing spatial inhomogeneities suggestive of incipient CP islanding\cite{Sacepe:NatPhys2011,Sacepe:NatComm2010,Mondal:PRL2011}.  Other tunneling measurements on insulating InO$_{x}$ films confirmed that the gap persists\cite{Sherman:PRL2012}. Earlier electron transport experiments on films near the SIT revealed a giant magnetoresistance (MR) peak larger than expected for any known single electron transport mechanism\cite{Gantmakher:JETP1998, Sambandamurthy:PRL2004, Steiner:PRL2005, Baturina:PRL2007}.  Moreover, strong evidence of locally phase coherent charge $2e$ transport in insulating InO$_{x}$ films was obtained using a technique described below\cite{Kopnov:PRL2012}.  It is now clear that this CPI phase appears in many film systems and its description by theories of the SIT has become pervasive. The most current theories of the disorder-driven SIT predict the formation of puddles of CPs that become phase incoherent with one another with increasing disorder\cite{Fisher:PRL1990, Ghosal:PRL1998, Bouadim:NatPhys2011, Dubi:Nature2007, Feigelman:APhys2010, Falco:PRL2010}.  This emergent granularity is predicted to appear even in uniformly disordered systems where the length scale of the disorder is much smaller than the correlated regions that form\cite{Bouadim:NatPhys2011}. 

On the other hand, early measurements on amorphous elemental films\cite{Valles:PRL1992, Valles:PRL1994, Chervenak:PRB1999, Liu:PRB1993, Parker:EPL2006} still lend support to theoretical models proposing that disorder-enhanced electron electron interactions drive the pair density to zero at the SIT\cite{Finkelstein:PhysicaB1994, Belitz:RMP1994}.  This fermionic mechanism has been used to explain the suppression of the superconducting transition temperature to zero with increasing disorder in elemental amorphous thin films\cite{Finkelstein:PhysicaB1994, Belitz:RMP1994}.  The most significant evidence for the fermionic mechanism is provided by early planar tunneling data on amorphous Bi films near their SIT\cite{Valles:PRL1992, Valles:PRL1994}.  These data indicated that the CP amplitude vanishes at the transition, which creates a problem for the bosonic picture. However, it is still possible that the bosonic scenario is true in general if the tunneling data do not reflect a vanishing CP density.  The data could be explained by the appearance of gapless superconductivity, or a very inhomogeneous distribution of energy gaps smearing the density of states. In this case, superconducting pair-correlations would always exist in insulators close to the SIT, albeit less dramatically in some cases. 

To study the evolution of the superconducting phase coherence through the disorder-driven SIT of uniform a-Bi films, we quench-deposited 1nm of Sb and then subsequent layers of Bi onto a Si substrate patterned with a nanohoneycomb hole array.   A large area nanometer-scale hole array was embedded into the Si using electrochemical processes and plasma etching.  A layer of AAO was grown into an Al film deposited on the Si using electrochemical anodization. Then a plasma etch was used to transfer the hole array into the Si, with the AAO as a mask.  Finally, the AAO layer was etched away with phosphoric acid.  This procedure produced an atomically flat Si substrate with holes in an approximately triangular array with an average hole radius of 23$\pm$12nm and spacing of 110$\pm$38nm (shown in the inset of Fig. 2a).  The patterned Si substrate was precoated with a 10nm-thick underlayer of Ge and Au/Ge contact pads at room temperature.  The Ge underlayer ensures a surface similar to a nearby reference film deposited on glass and the quench-deposited Sb underlayer makes the Bi wet the Si (or glass)/Ge substrate to produce uniform, amorphous films\cite{Ekinci:PRL1999,Ekinci:PRB1998}.  Film sheet resistances, $R_{\square}$, were measured using standard four-point AC and DC techniques in the linear response regime on a (1mm)$^{2}$ area of film. The sheet resistance of adjacent squares are similar to better than 15\% at 8K, indicating good uniformity of the film thickness on a mm scale. Additionally, the film's normal state  conductance grows linearly with thickness as expected for amorphous films\cite{Hollen:PRB2011}.

The disorder (thickness)-tuned SIT (d-SIT) of uniform, amorphous Bi films with an embedded hole array (Fig. 2a) closely resembles the d-SIT of uniform, amorphous Bi films on glass  (see \textit{e.g.} \textcite{Chervenak:PRB1999}). Thinner films are insulators as defined by $dR_{\square}/dT<0$.  With increasing thickness, the films become weaker insulators, developing a conductance, $G=1/R$, that decreases as $\Delta G = \alpha G_{00} \ln T$\cite{SuppMat}.  For film 6, the thickest film that does not exhibit superconducting fluctuations, $\alpha\simeq 1$.  This temperature dependence is consistent with corrections to the Drude conductance due to disorder enhanced electron-electron interaction and weak localization effects.  The prefactor, $\alpha$, is smaller than but of the same order as the prefactor found for films without holes, 1.2\cite{Chervenak:PRB1999}.  A slight increment to the film thickness produces film 7, whose $R(T)$ turns down at the lowest temperatures accessed.   We consider film 7, with $R_{N}=\mathrm{10.6k}\Omega$, to be very close to the critical point for the SIT because this downturn is a clear SC fluctuation effect\cite{Chervenak:PRB1999}.  This critical resistance is close to, but higher than, the $\simeq \mathrm{6.5 k}\Omega$ value for films on glass.  Subsequent evaporations produce films exhibiting sharp superconducting transitions.  The transition temperature for SC, as measured at the resistive midpoint, grows smoothly with thickness. Thus, these films with holes show the same qualitative behavior as films without holes.  

Moreover, the quantitative differences in $\alpha$ and the critical $R_{N}$ can be attributed to the influence of the tortuous geometry induced by the hole array. The conductance of a film with holes is smaller than an equivalent film without holes. The temperature dependence of the conductance is diminished by the same ratio.  The ratio of these conductances can be calculated using finite element analysis of Laplace's equation for the holey geometry provided the spacing between the holes exceeds the electronic mean free path and the length scales that affect the quantum corrections to the conductance.  For the average hole density and radii imposed by the Si substrate, this factor is 1.6.  It is comparable to the experimentally measured ratio indicating that the microscopic transport properties of Bi films on holey Si and glass\cite{Chervenak:PRB1999} or alumina substrates\cite{Liu:PRB1993} are nearly identical.  Also, it suggests that the critical normal state resistance for the SIT is dictated by the microscopic rather than the macroscopic properties.

Now we turn to the detection of locally phase coherent CPs in films closest to the d-SIT.  If CPs are present and phase coherent over a length exceeding the interhole spacing of $\sim$100nm, we expect to observe LP MR oscillations with a period corresponding to one flux quantum per hole ($H_\mathrm{M}$). For the hole array used to pattern these films (Fig. 2a, inset), $\mu_{0}H_\mathrm{M}= \Phi_{0}n_\mathrm{holes}=$0.22$\pm$0.02T, where $\Phi_{0}=h/2e$ is the flux quantum and $n_\mathrm{holes}$ is the areal density of holes.  As shown by Fig. 2b, superconducting films (films 8 and 9) display MR oscillations on a rising background at low fields.  The insulating films only display the monotonic rise.  At high fields, the MR saturates for both superconducting and insulating films. Even at the lowest temperatures accessed (T=130mK) there are no signs of the giant MR peak characteristic of the CPI state\cite{SuppMat, Sambandamurthy:PRL2004, Steiner:PRL2005, Baturina:PRL2007}. 

We also measured $R(T)$ at magnetic fields of 0, H$_\mathrm{M}$/2, and H$_\mathrm{M}$  (Fig. 3a) to obtain the temperature dependence of the local phase coherence signal, as well as a lower temperature investigation of the presence of LP MR oscillations.  Defining the oscillation amplitude as,
 \begin{equation}
A=R(\mathrm{H}_\mathrm{M}/2,T)-R_\mathrm{mid}(T);
\end{equation}
\begin{equation}
R_\mathrm{mid}(T)\equiv\frac{1}{2} \left( R(0,T)+R(\mathrm{H}_\mathrm{M},T)\right).
 \end{equation}
Fig. 3b shows that $A$ (normalized by $R_\mathrm{mid}$) rapidly grows from zero as $T$ drops below the mean field transition temperature of the superconducting films.  For the film closest to the d-SIT, there is a hint that an oscillation due to Cooper pairing begins to emerge at the lowest temperatures.  It is notable that this hint appears as the film resistance starts to drop. 
 For films where no such drop in $R_{\square}$ is observed, there is no evidence of local phase coherence. 
 
 These results indicate that uniformly thick a-Bi films do not undergo a disorder-tuned superconductor to CPI transition.  The phase coherence length appears to collapse to zero at this SIT.  While it is true that this behavior could indicate a first order transition at which CPs abruptly localize, it seems more likely that this SIT is fermionic.  Previous measurements of the tunneling density of states suggested that the superconducting energy gap decreases continuously toward zero on approaching this SIT\cite{Valles:PRL1992, Valles:PRL1994}.  Since the gap is proportional to the CP density, this behavior suggested that the order parameter amplitude goes to zero at the SIT.  The present results support this interpretation.  The suppression of the order parameter amplitude to zero leaves no room for phase coherence of the order parameter over any scale.  
 
To summarize, the disorder-tuned SIT of amorphous elemental films of Bi prepared with a uniform film thickness involves an insulator phase that differs fundamentally from the CPI  phase that appears for non-uniform film thickness.  This interpretation implies that there are at least two distinct classes of disorder-driven SIT; a conclusion that conflicts with recent predictions that the interplay of disorder and quantum fluctuation effects leaves $\Delta$ finite at the critical point\cite{Bouadim:NatPhys2011, Dubi:Nature2007, Feigelman:APhys2010}.  The fact that amorphous elemental films can exhibit either SIT depending on their morphology underlines the limitations of describing disorder using a single parameter like the sheet resistance.\cite{Beloborodov:PRL2004}
Perhaps the data indicate that repulsive Coulomb interactions, which are known to reduce $\Delta$ with disorder\cite{Finkelstein:PhysicaB1994, Belitz:RMP1994}, play an essential role that varies with film structure. 
The resolution of this issue is important for our fundamental understanding of the SIT as well as the interpretation of qualitatively similar SITs, such as those of the high T$_{c}$ superconductors\cite{Bollinger:Nature2011, Beschoten:PRL1996, Leng:PRL2011}.

We gratefully acknowledge helpful conversations with N. Trivedi, T. Baturina, R. Barber, and M.D. Stewart, Jr.  This work was supported by the NSF through Grants No. DMR-0605797 and No. DMR-0907357, by the AFRL, the ONR, the AFOSR, and the WCU program at SNU, Korea. We are also grateful for the support of AAUW.

 \begin{figure}
\begin{center}
\includegraphics[width=0.5\columnwidth,keepaspectratio]{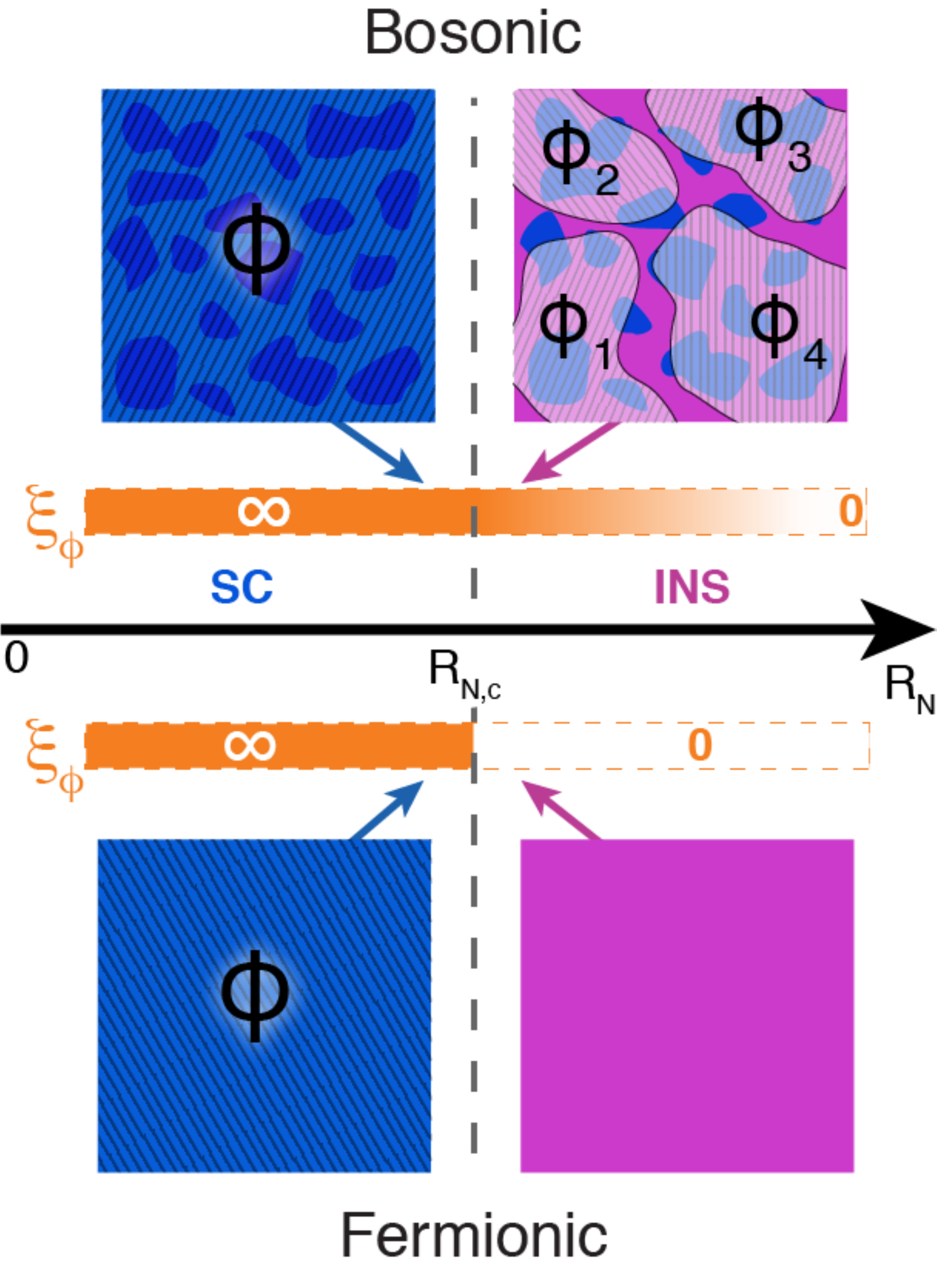}
\caption{\textbf{Disorder-tuned superconductor-insulator transition types} Illustrations of disorder-driven SITs involving the emergence of inhomogeneities that lead to Cooper pair islanding (bosonic SIT) and pair breaking in homogeneous films (fermionic SIT).  The superconducting (SC) to insulating (INS) film transitions occur with increasing normal state sheet resistance, $R_{N}$, at a critical value $R_{Nc}$. Blue regions are paired, and pink regions are unpaired.  The orange bar depicts the expected Cooper pair phase coherence length, $\xi_{\phi}$, which sets the size of the phase coherent regions in the bosonic insulator.  The parallel stripes represent the phase angle of the superconducting wave function, denoted as $\phi$.
\label{cap:fig1}}
\end{center}
\end{figure}

\begin{figure}
\begin{center}
\includegraphics[width=1\columnwidth,keepaspectratio]{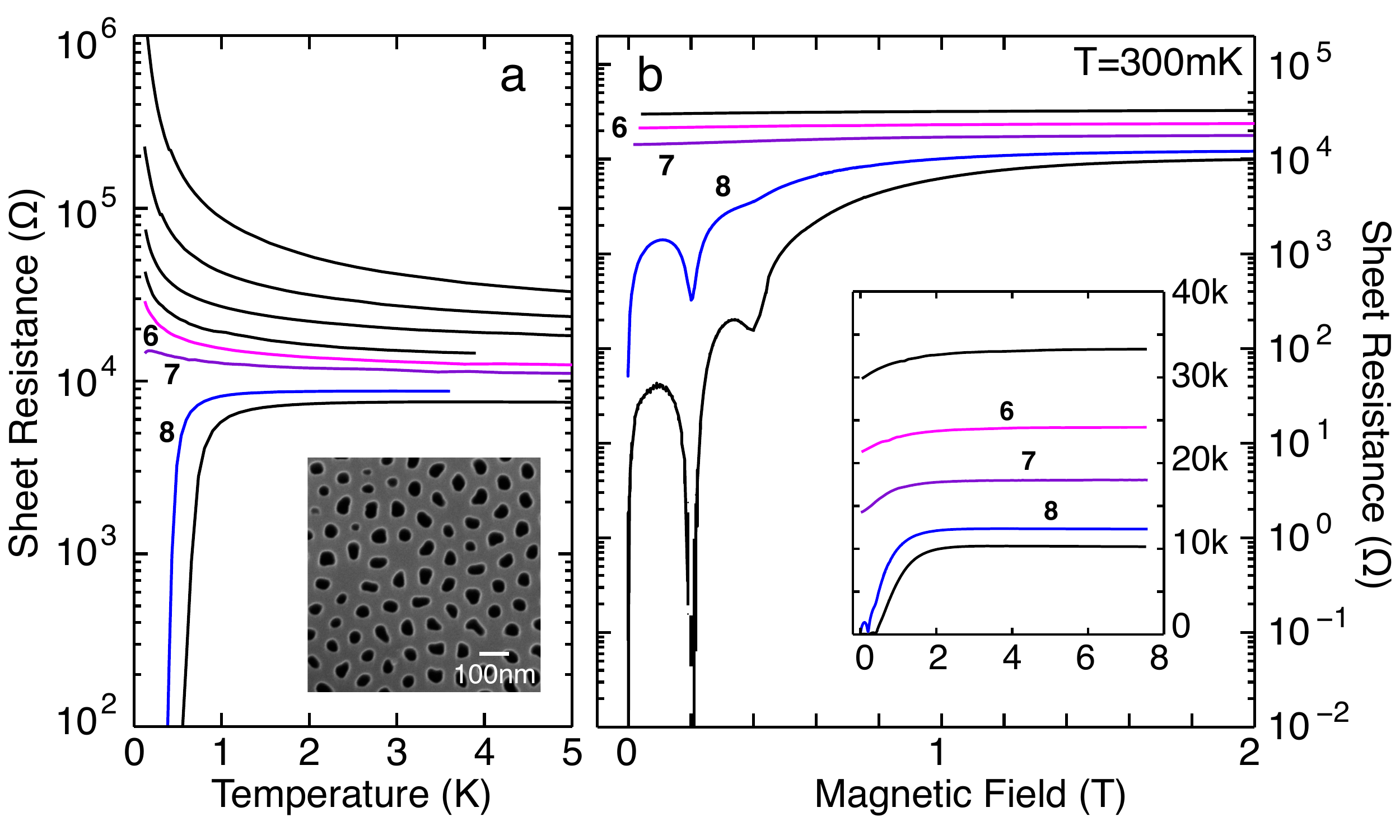}
\caption{\textbf{Global and local phase coherence through the SIT} a) Sheet resistance with temperature at H=0T for a series of uniform, amorphous Bi films quench condensed onto a Si substrate with a nanometer-scale array of holes (inset). The hole array has an average hole radius of 23$\pm$12nm and spacing of 110$\pm$38nm.  The film thicknesses range from 0.47 to 0.71nm. b) Magnetoresistance (MR) (for field applied perpendicular to film plane) at T=300mK for films 5-9 in (a) on a log scale. The dips spaced by $H_\mathrm{M}=$0.21T indicate the presence of CPs.  They appear for (superconducting) films 8 and 9 only. Inset: MR data on a linear scale for films 5-9 of (a) that shows the absence of an MR peak\cite{Sambandamurthy:PRL2004, Steiner:PRL2005, Baturina:PRL2007} at high fields. 
\label{cap:fig2}}
\end{center}
\end{figure}

\begin{figure}
\begin{center}
\includegraphics[width=1\columnwidth,keepaspectratio]{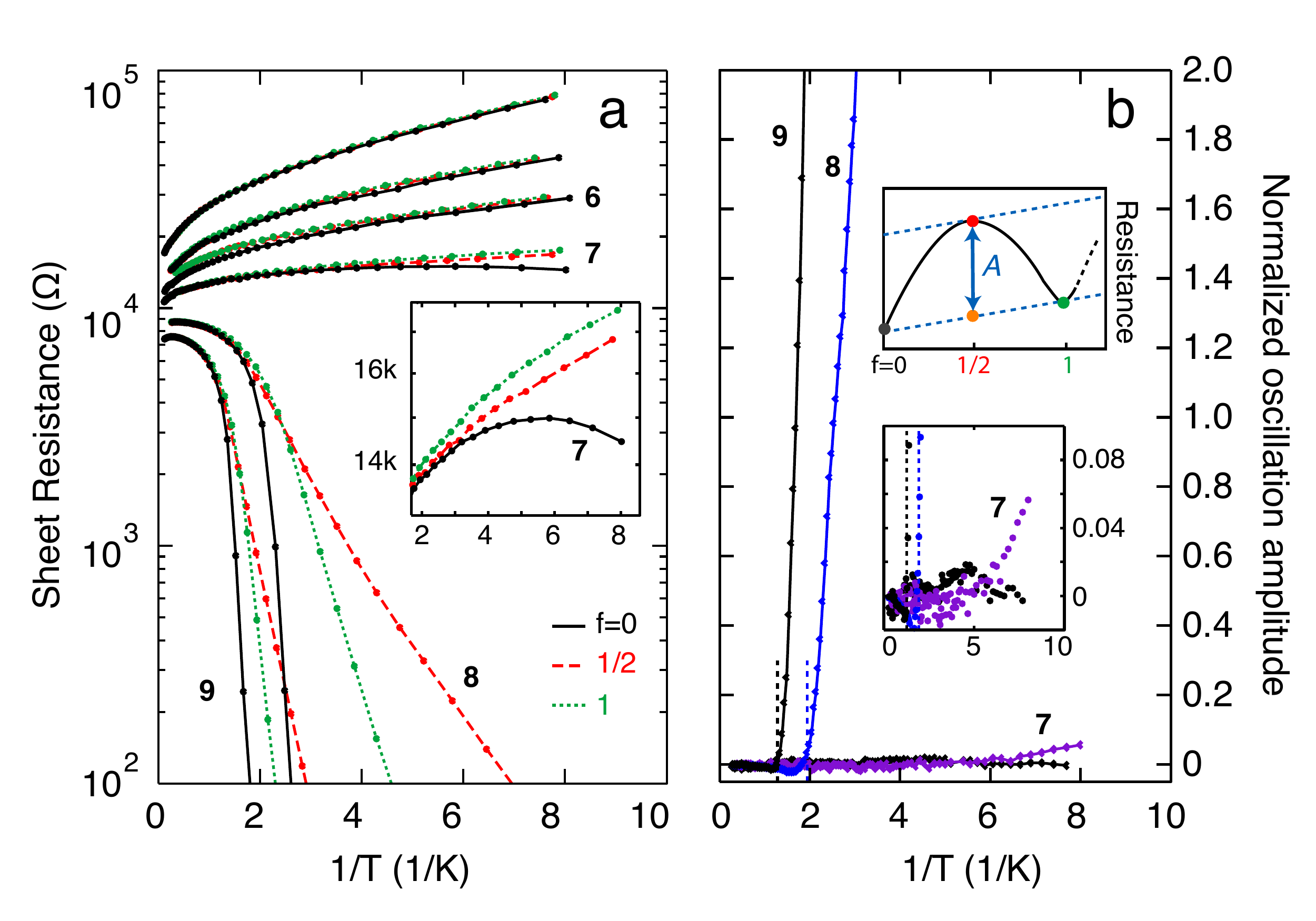}
\caption{\textbf{Temperature dependence of local phase coherence signal} a) Sheet resistance with temperature on an Arrhenius scale for the series of films in Fig. 1a at three perpendicular magnetic field values: $\mu_{0}H_{\perp}=$0 (solid line), $\mu_{0}H_\mathrm{M}/2$=0.105T (dashed red line), and $\mu_{0}H_\mathrm{M}$=0.21T (dotted green line) ($f\equiv H/H_\mathrm{M}$). Inset: magnified view showing the drop in resistance at the lowest temperatures for film 7 at H=0T. b) Normalized oscillation amplitude, $A/R_\mathrm{mid}$ (defined in upper inset and equations (1) and (2)), versus inverse temperature for films 5 and 7-9. Vertical dashed lines mark mean-field $T_{c}s$ for superconducting films: 0.52 and 0.78K for films 8 and 9 respectively. Lower inset: magnified view of the growth of oscillation amplitude at the lowest temperatures in film 7.
\label{cap:fig3}}
\end{center}
\end{figure}

\clearpage
\bibliographystyle{apsrev}
\bibliography{Manuscript}

\end{document}